\documentclass[aps,pra,showpacs,superscriptaddress,twocolumn]{revtex4}
\usepackage{amssymb}
\usepackage{mathrsfs}
\usepackage{amsmath}
\usepackage{graphicx}
\usepackage{multirow}
\usepackage{hhline}

\begin{document}

\begin{titlepage}
\title
{The transition from quantum Zeno to anti-Zeno effects for a qubit
in a cavity by varying the cavity frequency}
\author{Xiufeng Cao\footnote{Email: xfcao@xmu.edu.cn \\
Fax: 080-0592-2189426}} \affiliation{Department of Physics and
Institute of Theoretical Physics and Astrophysics, Xiamen
University, Xiamen, 361005,China} \affiliation{Advanced Science
Institute, RIKEN, Wako-shi 351-0198, Japan}
\author{Qing Ai}
\affiliation{Advanced Science Institute, RIKEN, Wako-shi 351-0198,
Japan} \affiliation{Institute of Theoretical Physics, Chinese
Academy of Sciences, Beijing 100190, People's Republic of China}
\author{C. P. Sun}
\affiliation{Advanced Science Institute, RIKEN, Wako-shi 351-0198,
Japan} \affiliation{Institute of Theoretical Physics, Chinese
Academy of Sciences, Beijing 100190, People's Republic of China}
\author{Franco Nori}
\affiliation{Advanced Science Institute, RIKEN, Wako-shi 351-0198,
Japan} \affiliation{Physics Department, The University of Michigan,
Ann Arbor, Michigan 48109-1040, USA}
\date{\today }

\begin{abstract}
We propose a strategy to demonstrate the transition from the quantum
Zeno effect (QZE) to the anti-Zeno effect (AZE) using a
superconducting qubit coupled to a transmission line cavity, by
varying the central frequency of the cavity mode. Our results are
obtained without the rotating wave approximation (RWA), and the
initial state (a dressed state) is easy to prepare. Moreover, we
find that in the presence of both qubit's intrinsic bath and the
cavity bath, the emergence of the QZE and the AZE behaviors relies
not only on the match between the qubit energy level spacing and the
central frequency of the cavity mode, but also on the coupling
strength
between the qubit and the cavity mode. 


\end{abstract}

\maketitle

\end{titlepage}

\section{Introduction}

The QZE predicts that the decay rate of a system can be slowed down by
measuring it frequently enough \cite%
{pra-41-2295,pra-42-5720,pra-44-1466,pascazio}. However some systems are
predicted to have an enhancement of the decay due to the frequent
measurements, namely the AZE or inverse Zeno effect \cite{nature-405-546,p,q}%
. The QZE and AZE have been observed in an unstable system \cite%
{prl-87-040402}.

Recently, the QZE-AZE crossover in quantum Brownian motion model was
investigated \cite{prl-97-130402}, where a system of damped harmonic
oscillator interacts with a bosonic reservoir in thermal equilibrium. It was
found \cite{prl-97-130402} that controlling the system-environment coupling
by an artificially-controllable engineered environment (e.g., \cite%
{nature-407-57,nature-403-269}) would allow one to monitor the transition
from the QZE to the AZE dynamics. The QZE and AZE of a nanomechanical
resonator measured by a quantum point contact detector (non-equilibrium
fermionic reservoir) also was studied \cite{prb-81-115307}. Therefore,
modulating the system and reservoir parameters can induce the QZE-AZE
crossover.

In cavity QED, the coupling between the qubit and the cavity, in which the
electromagnetic field modes are concentrated around the cavity resonant
frequency, depends on the cavity frequency. For an excited qubit located in
a cavity, the cavity mode is the dominant one available for the qubit to
emit photons. If the qubit energy level spacing is resonant with the cavity
mode, the rate of decay into the particular cavity mode is enhanced.
Otherwise, it is inhibited. Therefore, one may manipulate the qubit decay
rate by varying the central frequency of the cavity mode in or off resonance
with the qubit level energy spacing. The variation of the qubit decay \cite%
{pra-81-062131,pra-54-R3750} in the cavity is an increasingly important
topic for experimental and theoretical studies \cite%
{prl-85-2272,prl-101-180402,pra-81-062131}.

In this paper, we propose to modulate the qubit's decay rate in cavity QED
by the QZE, which means invoke the frequent measurements in the qubit and
achieve the transition between QZE and AZE. We study a model of a qubit in a
cavity, and investigate the occurrence of either the QZE or AZE by varying
the cavity central frequency. We insert frequent projection measurements in
the qubit decay process and find that the normalized decay rate depends on
whether the central frequency of the cavity mode is in resonance with the
qubit energy level spacing or not. In the resonant case, the normalized
decay rate is lower than 1, so the QZE of the qubit occurs. However, when
the cavity mode is detuned from the energy level spacing of the qubit, the
normalized decay rate is larger than 1 and the qubit exhibits AZE. The
variation from the QZE to the AZE, by varying the central frequency of the
cavity mode, should help distinguishing these two kinds of effects.
Moreover, we consider the case when both the qubit's intrinsic bath and the
cavity bath are simultaneously present. And find the dependence of the
behaviors (the QZE and the AZE) on the coupling strength of the qubit-cavity
and the cavity central frequency.

The QZE-AZE crossover may be achieved in a superconducting qubit coupled to
a transmission line cavity \cite%
{jqyou,prl-103-147003,pra-81-042304,pra-69-062320}. This is because there
are two physical mechanisms to tune the resonant frequency of the
transmissionline resonator. One method is to change the boundary condition
of the electromagnetic field in the transmission line \cite%
{prb-74-224506,apl-92-203905,apl-93-042510}, as shown in the Fig.~\ref%
{Fig0ab}(a). Another method is to construct a transmission line resonator by
using a series of magnetic-flux biased SQUIDs, as shown in the Fig.~\ref%
{Fig0ab}(b). Because the effective inductor of a magnetic-flux-biased SQUID
can be tuned by changing the applied magnetic flux \cite%
{nature-4-929,apl-91-083509}, the inductance per unit length of the SQUID
array is controllable.
\begin{figure}[th]
\includegraphics[width=3.0in,clip]{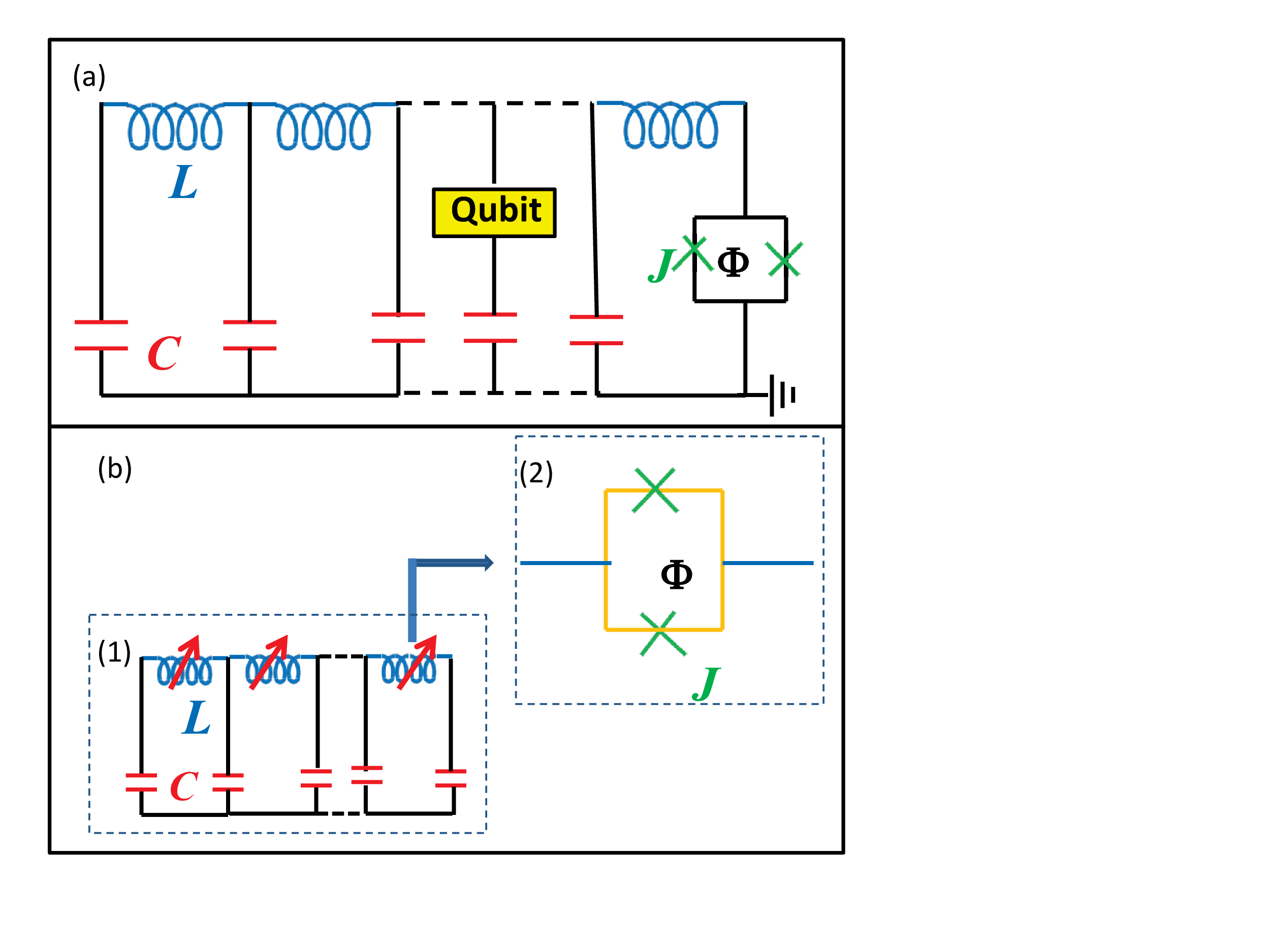}
\caption{(Color online) (a) Superconducting circuit model of a
frequency-tunable transmission line resonator, which is archived by changing
the boundary condition, coupled with a qubit. (b) Superconducting circuit
model (1) of the effective tunable inductors, which are consisted of a
series array of SQUIDs (2).}
\label{Fig0ab}
\end{figure}

\section{Hamiltonian of a qubit in a cavity beyond the rotating wave
approximation}

\begin{figure}[th]
\includegraphics[width=3.2in,clip]{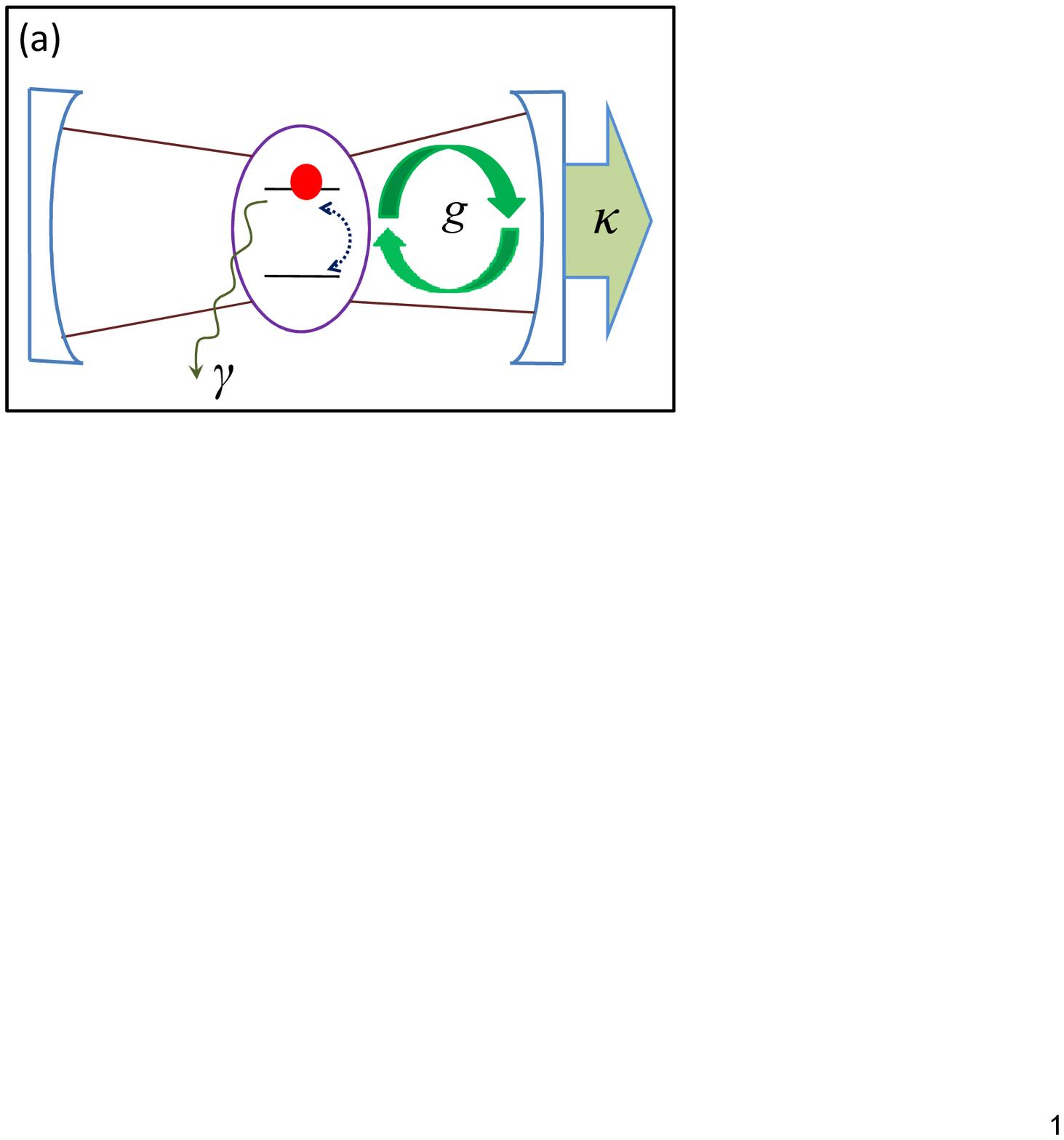}
\par
\includegraphics[width=3.1in,clip]{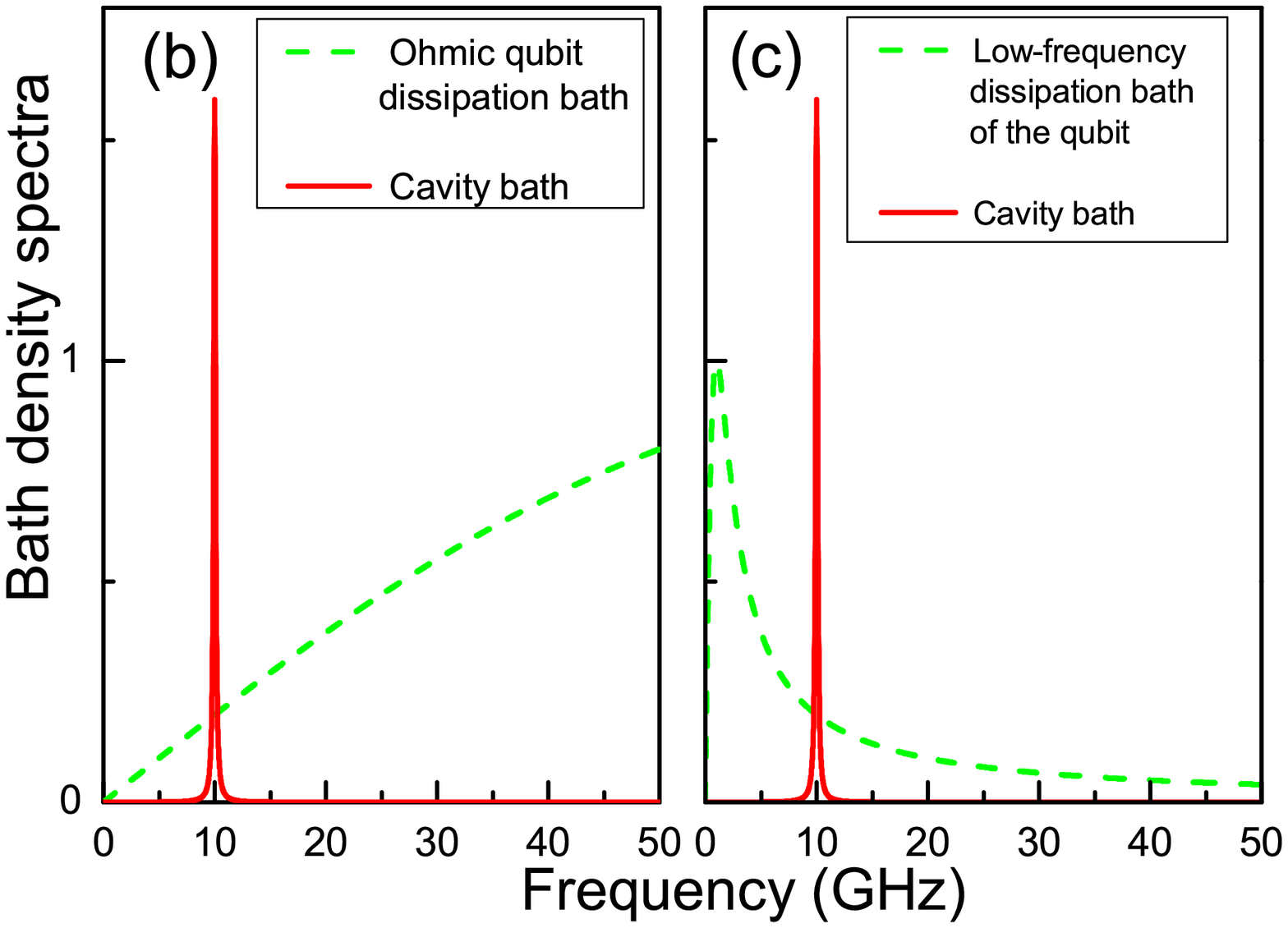}
\caption{(Color online) (a) Sketch of a qubit with the spontaneous
dissipation rate $\protect\gamma $ coupled to a cavity with the loss rate $%
\protect\kappa $ via a coupling strength $g.$ (b) and (c) schematically show
the bath density spectrum of the qubit environment: (b) the Ohmic qubit's
intrinsic bath (green dashed) and the Lorentzian cavity bath (red solid),
(c) the low-frequency qubit's intrinsic bath (green dashed) and the
Lorentzian cavity bath (red solid).}
\label{Fig1}
\end{figure}

Including the qubit dissipation environment, the Hamiltonian of a qubit in a
lossy cavity can be written as%
\begin{eqnarray}
H &=&\frac{1}{2}\Delta \sigma _{z}+\sum_{k}\omega _{k,1}b_{k}^{\dagger
}b_{k}+\sum_{k}f_{k}(b_{k}^{\dagger }+b_{k})\sigma _{x}  \notag \\
&&+\sum_{k}\omega _{k,2}a_{k}^{\dagger }a_{k}+\sum_{k}g_{k}(a_{k}^{\dagger
}+a_{k})\sigma _{x}.  \label{e1}
\end{eqnarray}%
The Pauli operators, $\sigma _{z}$ and $\sigma _{x},$ describe the qubit
level energy spacing and tunneling. The operators $b_{k}$ and $%
b_{k}^{\dagger }$ are the annihilation and creation operators characterizing
the qubit's intrinsic bath with frequencies $\omega _{k,1}$. The lossy
cavity is modeled as a collection of harmonic oscillators with frequencies $%
\omega _{k,2},$ with the creation operators $a_{k}^{\dagger }$ and the
annihilation\ operators $a_{k}$. Figure \ref{Fig1}(a) schematically shows
the model considered here.
Notice that \textit{no }RWA is invoked in the Hamiltonian $H$ and thus it
can not be diagonalized exactly.

Now let us solve the Schr{\"{o}}dinger equation of the Hamiltonian~(\ref{e1}%
). We take the \textit{anti-rotating} terms into account, which guarantees
that our discussions extend to the off-resonant regime and also the case
when there is a strong qubit-cavity interaction. Due to the anti-rotating
terms, we apply a unitary transformation to the Hamiltonian $H$,%
\begin{equation}
H^{\prime }=\exp (S)H\exp \left( -S\right) ,  \label{u}
\end{equation}%
with%
\begin{equation}
S=\sum_{k}\left[ \frac{f_{k}}{\omega _{k,1}}\xi _{k,1}(b_{k}^{\dagger
}-b_{k})+\frac{g_{k}}{\omega _{k,2}}\xi _{k,2}(a_{k}^{\dagger }-a_{k})\right]
\sigma _{x}.
\end{equation}%
Here, the $k$-dependent variables%
\begin{equation}
\xi _{k,1}=\omega _{k,1}/(\omega _{k,1}+\eta _{1}\;\Delta ),
\end{equation}%
and%
\begin{equation}
\xi _{k,2}=\omega _{k,2}/(\omega _{k,2}+\eta _{2}\;\Delta ),
\end{equation}%
are introduced in the transformation. The transformed Hamiltonian $H^{\prime
}$ can be written as%
\begin{eqnarray}
H^{\prime } &\approx &\frac{1}{2}\eta \;\Delta \sigma _{z}+\sum_{k}\omega
_{k,1}b_{k}^{\dagger }b_{k}+\sum_{k}\omega _{k,2}a_{k}^{\dagger }a_{k}
\notag \\
&&+\sum_{k}V_{k,1}(b_{k}^{\dagger }\sigma _{-}+b_{k}\sigma _{+})  \notag \\
&&+\sum_{k}V_{k,2}(a_{k}^{\dagger }\sigma _{-}+a_{k}\sigma _{+}),  \label{E4}
\end{eqnarray}%
with $\sigma _{\pm }=\left( \sigma _{x}\pm i\sigma _{y}\right) /2$ and%
\begin{equation}
\eta =\eta _{1} \: \eta _{2}.
\end{equation}%
Then, the qubit energy-level-spacing $\Delta $ is renormalized to $\eta
\;\Delta $ because of its coupling to the qubit's intrinsic bath and the
cavity bath. These factors $\eta _{1}$ and $\eta _{2},$ are respectively
denoted by%
\begin{eqnarray}
\eta _{1} &=&\exp \left( -\sum_{k}2f_{k}^{2}\xi _{k,1}^{2}{}/\omega
_{k,1}^{2}\right) , \\
\eta _{2} &=&\exp \left( -\sum_{k}2g_{k}^{2}\xi _{k,2}^{2}{}/\omega
_{k,2}^{2}\right) .
\end{eqnarray}%
The coupling constants $f_{k}$ and $g_{k},$ of the qubit-environment
interaction are also renormalized. The renormalized factors are respectively
denoted by%
\begin{eqnarray}
V_{k,1} &=&2\eta _{1}\;\Delta \;f_{k}/\left( \omega _{k,1}+\eta _{1}\;\Delta
\right) , \\
V_{k,2} &=&2\eta _{2}\;\Delta \;g_{k}/\left( \omega _{k,2}+\eta _{2}\;\Delta
\right) ,
\end{eqnarray}%
owing to the anti-rotating coupling terms. In Eq.~(\ref{E4}), we drop the
higher-order terms, which include the induced effect of the two baths by the
coupling to the same qubit $\mathcal{O}$$\left( f_{k}\cdot g_{k}\right) $,
whose contributions to the physical quantities are of the order $\mathcal{O}$%
$\left( g_{k}^{4}\right) $ $\left[ \text{or }\mathcal{O}\left(
f_{k}^{4}\right) ,\text{ or }\mathcal{O}\left( f_{k}^{2}g_{k}^{2}\right) %
\right] $ and higher.

\section{Equation of motion of a qubit in a cavity beyond the rotating wave
approximation}

Below, we will solve the equation of motion of the wave function, beyond the
RWA, in the transformed Hamiltonian $H^{\prime }$ in Eq.~(\ref{E4}). Since
the total excitation number operator%
\begin{equation}
N=\sum_{k}\left( a_{k}^{\dagger }a_{k}+b_{k}^{\dagger }b_{k}\right) +\left(
1+\sigma _{z}\right) /2,
\end{equation}%
of the dissipative qubit-cavity system is a conserved observable, i.e., $%
\left[ N,H^{\prime }\right] =0,$ it is reasonable to restrict our discussion
to the single-particle excitation subspace. A general state in this subspace
can be written as
\begin{equation}
\left\vert \Phi (t)\right\rangle =\chi (t)\left\vert \uparrow \right\rangle
\left\vert \{0_{k}\;0_{k}\}\right\rangle +\sum_{k,i}\beta
_{k,i}(t)\left\vert \downarrow \right\rangle \left\vert \{0_{k,\overline{i}%
}\;1_{k,i}\}\right\rangle ,
\end{equation}%
where $\left\vert \uparrow \right\rangle $ and $\left\vert \downarrow
\right\rangle $ are the eigenstates of $\sigma _{z}$ ($\sigma _{z}\left\vert
\uparrow \right\rangle =\left\vert \uparrow \right\rangle $ and $\sigma
_{z}\left\vert \downarrow \right\rangle =-\left\vert \downarrow
\right\rangle $), the state $\left\vert \{0_{k,\overline{i}%
}\;1_{k,i}\}\right\rangle $ ($i$ can be $1,2$) means that either the cavity
bath or the qubit's intrinsic bath has one quantum excitation. Substituting $%
\left\vert \Phi (t)\right\rangle $ into the Schr{\"{o}}dinger equation, we
have
\begin{equation}
i\frac{d\chi (t)}{dt}=\frac{\eta \;\Delta }{2}\chi
(t)+\sum_{k,i}V_{k,i}\;\beta _{k,i}(t),  \label{ee1}
\end{equation}%
\begin{equation}
i\frac{d\beta _{k,i}(t)}{dt}=\left( \omega _{k,i}-\frac{\eta \;\Delta }{2}%
\right) \beta _{k,i}(t)+\sum_{k,i}V_{k,i}\;\chi (t).  \label{ee2}
\end{equation}%
Applying the transformation
\begin{eqnarray}
\chi (t)\! &\!=\!&\!\widetilde{\chi }(t)\exp \left( -i\frac{\eta \;\Delta }{2%
}t\right) ,  \label{e3} \\
\beta _{k,i}(t)\! &\!=\!&\!\widetilde{\beta }_{k,i}(t)\exp \left[ -i\left(
\omega _{k,i}-\frac{\eta \;\Delta }{2}\right) t\right] ,  \label{e4}
\end{eqnarray}%
Eqs.~(\ref{ee1}) and (\ref{ee2}) is simplified as
\begin{equation}
\frac{d\widetilde{\chi }(t)}{dt}\!=\!-i\sum_{k,i}V_{k,i}\;\widetilde{\beta }%
_{k,i}(t)\exp \left[ -i(\omega _{k,i}-\eta \;\Delta )t\right] ,  \label{e5}
\end{equation}%
\begin{equation}
\frac{d\widetilde{\beta }_{k,i}(t)}{dt}=-iV_{k,i}\;\widetilde{\chi }(t)\exp %
\left[ i(\omega _{k,i}-\eta \;\Delta )t\right] .  \label{e6}
\end{equation}%
Integrating Eq.~(\ref{e6}) and substituting it into Eq.~(\ref{e5}), we
obtain
\begin{equation}
\frac{d\widetilde{\chi }(t)}{dt}=-\int\limits_{0}^{t}\sum_{k,i}V_{k,i}^{2}%
\exp \left[ -i(\omega _{k,i}-\eta \;\Delta )(t-t^{\prime })\right] \;%
\widetilde{\chi }(t^{\prime })\;dt^{\prime }.  \label{e7}
\end{equation}%
This integro-differential equation can be solved exactly by a Laplace
transformation,
\begin{equation}
\overline{\widetilde{\chi }(p)}=\frac{\widetilde{\chi }(0)}{%
p+\sum_{k,i}V_{k,i}^{2}/\left[ p-i(\eta \;\Delta -\omega _{k,i})\right] },
\end{equation}%
with%
\begin{equation}
\overline{\widetilde{\chi }(p)}=\int \widetilde{\chi }(t)\exp (-pt)dt.
\end{equation}%
Inversing of the Laplace transformation, we obtain the amplitude in the
excited-state%
\begin{equation}
\widetilde{\chi }(t)=\frac{1}{2\pi i}\int_{\sigma -i\infty }^{\sigma
+i\infty }\frac{\widetilde{\chi }(0)\exp (pt)}{p+\sum_{k,i}V_{k,i}^{2}/\left[
p-i(\eta \;\Delta -\omega _{k,i})\right] }dp
\end{equation}%
Then replace $p$ to $i\omega +0^{+},$%
\begin{equation}
\widetilde{\chi }(t)=\frac{1}{2\pi i}\int_{-\infty }^{\infty }\frac{%
\widetilde{\chi }(0)\exp (i\omega t)}{\omega -\sum_{k,i}V_{k,i}^{2}/\left[
(\omega +\eta \;\Delta )-\omega _{k,i}-i0^{+}\right] }d\omega
\end{equation}%
Denote $R(\omega )$ and $\Gamma (\omega )$ as the real and imaginary parts
of the summation term $\sum_{k,i}V_{k,i}^{2}/(\omega -\omega _{k,i}-i0^{+}),$
then%
\begin{eqnarray}
R(\omega ) &=&\wp \sum_{k,i}V_{k,i}^{2}/(\omega -\omega _{k,i}-i0^{+}) \\
\Gamma (\omega ) &=&\pi \sum_{k,i}V_{k,i}^{2}\delta (\omega -\omega
_{k,i}-i0^{+})
\end{eqnarray}%
where $\wp $ is the Cauchy principal value. Applying the pole approximation,%
\begin{equation}
\widetilde{\chi }(t)=\widetilde{\chi }(0)\sum_{j}\exp (i\omega
_{j}t)Q_{j}(\omega _{j})
\end{equation}%
where $\omega _{j}$ corresponds to the singularity of the quantity $%
\overline{\widetilde{\chi }(p)}$ and $Q_{j}(\omega _{j})$ is the normalized
factor.

Before doing further calculations, let us now focus on the initial state of
the system $\widetilde{\chi }(0)$, since different initial states may result
in distinct predictions about the QZE and the AZE \cite{PRL-101-200404,
pra-ai}. Indeed, these two effects can strongly depend on the initial
conditions. Through\ the unitary transformation in Eq.~(\ref{u}), the
Hamiltonian (\ref{e1}), which contains the anti-rotating terms, is reduced
to $H^{\prime }$ in Eq.~(\ref{E4}), which has the similar form of the
Hamiltonian under the RWA, with the parameters renormalized. Under energy
conservation, the ground state of $H^{\prime }$ is $\left\vert g^{\prime
}\right\rangle =\left\vert \downarrow \right\rangle \left\vert
\{0_{k}\;0_{k}\}\right\rangle $ and the corresponding ground-state energy is
$-\eta \;\Delta /2$. Therefore, through inversing the unitary
transformation, we obtain the ground state of the original Hamiltonian $H$
as $\left\vert g\right\rangle =\exp [-S]\left\vert \downarrow \right\rangle
\left\vert \{0_{k}\;0_{k}\}\right\rangle ,$ which is a dressed state of the
qubit and its environment due to the anti-rotating terms \cite%
{prl-103-147003,pra-80-053810}. In this paper, we choose the excited state $%
\exp [-S]\left\vert \uparrow \right\rangle \left\vert
\{0_{k}\;0_{k}\}\right\rangle $ as the initial state, which can be achieved
by acting the operator $\sigma _{x}$ on the ground state,%
\begin{equation}
\left\vert \psi (0)\right\rangle =\sigma _{x}\left\vert g\right\rangle =\exp
[-S]\left\vert \uparrow \right\rangle \left\vert
\{0_{k}\;0_{k}\}\right\rangle .
\end{equation}%
Thus, the initial state after the transformation is $\left\vert \psi
^{\prime }(0)\right\rangle =\left\vert \uparrow \right\rangle \left\vert
\{0_{k}\;0_{k}\}\right\rangle $, correspondingly the excited-state
probability amplitude $\chi (0)=1.$

To obtain the final result, we need the knowledge of the interacting spectra
of the qubit's intrinsic bath and also the cavity bath. From the quasi-mode
approach, the qubit-cavity coupling density spectrum is a Lorentzian density
spectrum \cite{pra-71-032302,a}
\begin{equation}
J_{\mathrm{cav}}(\omega )=\sum g_{k,2}^{2}\delta (\omega -\omega _{k})=\frac{%
g^{2}\;\lambda }{\pi \lbrack (\omega -\omega _{\mathrm{cav}})^{2}+\lambda
^{2}]},
\end{equation}%
where $g$ is the coupling constant between the cavity and the qubit, $\omega
_{\mathrm{cav}}$ the central frequency of the cavity mode, and $\lambda $ is
the frequency width of the cavity bath density spectrum and is related to
the cavity bath correlation time. The physical quantity $\omega _{\mathrm{cav%
}}/\lambda $ denotes the quality factor $Q$ of the cavity.

Experiments in some superconducting qubits indicate that the noise chiefly
comes from the low-frequency region. The density spectrum of the
low-frequency bath can be approximately written as
\begin{equation}
J_{\mathrm{qb}}^{\mathrm{low}}(\omega )=\sum_{k}g_{k,2}^{2}\delta (\omega
-\omega _{k,1})=\frac{2\;\alpha _{\mathrm{low}}\;\omega }{\left( \omega
/\Delta \right) ^{2}+\left( \omega _{\mathrm{low}}/\Delta \right) ^{2}}.
\label{E2}
\end{equation}%
where $\omega _{\mathrm{low}}$ is an energy lower than the qubit energy
spacing $\Delta $, and $\alpha _{\mathrm{low}}$ a dimensionless coupling
strength between the qubit and the intrinsic bath. In semiconductor quantum
dot qubits, the qubit spontaneous dissipation bath, mainly the phonon bath,
is usually described by an Ohmic density spectrum. Thus, the density
spectrum $J_{\mathrm{qb}}(\omega )$ of the Ohmic bath with Drude cutoff can
be given as%
\begin{equation}
J_{\mathrm{qb}}^{\mathrm{Ohm}}(\omega )=\sum_{k}g_{k,1}^{2}\delta (\omega
-\omega _{k,1})=\frac{2\;\alpha _{\mathrm{Ohm}}\;\omega }{1+\left( \omega
/\omega _{\mathrm{Ohm}}\right) ^{2}},
\end{equation}%
where $\omega _{\mathrm{Ohm}}$ is the high-frequency cutoff, which is
typically assumed to be larger than the qubit energy level spacing, and $%
\alpha _{\mathrm{Ohm}}$ is the dimensionless coupling strength.

So we consider three kinds of interacting density spectra: Lorentzian cavity
bath, low-frequency qubit's intrinsic bath and Ohmic qubit's intrinsic bath,
and present a sketch of the density spectra of the qubit environment in Fig. %
\ref{Fig1}(b, c): showing the same cavity bath and different qubit's
intrinsic baths (a low-frequency bath in (b) and an Ohmic bath in (c)).
\begin{figure}[tbp]
\includegraphics[width=3.2in,clip]{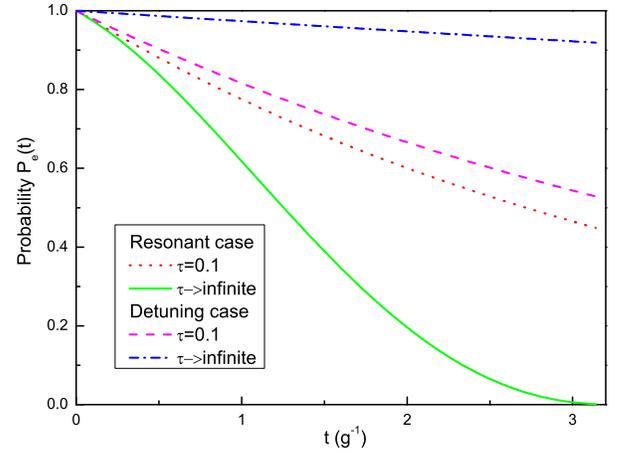}
\caption{(Color online) Time dependence of the probability for the qubit at
its excited state. In the resonant case, the parameters are $\protect\omega %
_{\mathrm{cav}}=\Delta =100~g$ and $\protect\tau =0.1~g^{-1}$. In the
detuning case, the cavity mode frequency is varied to $\protect\omega _{%
\mathrm{cav}}=80~g$. Note that the successive measurements slow down the
decay rate of excited state in the resonant case, which is the QZE. While in
the detuning case, the measurements speed up the qubit decay rate, which is
the AZE.}
\label{Fig22}
\end{figure}

Before illustrate our results, let us recall the standard master equation of
a qubit coupled to a single-mode cavity under the RWA and Markov
approximation \cite{pra-40-5516}
\begin{eqnarray}
\dot{\rho} &=&-i\left[ H_{\mathrm{RWA}},\rho \right] +\gamma \left( 2\sigma
_{-}\rho \sigma _{+}-\sigma _{+}\sigma _{-}\rho -\rho \sigma _{+}\sigma
_{-}\right)  \notag \\
&&+\kappa \left( 2a\rho a^{\dagger }-a^{\dagger }a\rho -\rho a^{\dagger
}a\right) ,
\end{eqnarray}%
where $H_{\mathrm{RWA}}=g(\sigma _{-}a^{\dagger }+\sigma _{+}a),$ $g$ is the
qubit-cavity coupling strength, $a^{\dagger }$ and $a$ are the creation and
annihilation operators for the single-mode cavity. The two parameters $%
\kappa $ and $\gamma $ correspond to the decay rates induced by the two
baths: the qubit's intrinsic bath and the cavity bath, respectively. Then
the survival probability of the qubit in the excited state is approximately
\cite{pra-40-5516}%
\begin{equation}
P_{\mathrm{e}}(t)=\left\vert \chi (t)\right\vert ^{2}\;=\cos \left(
gt\right) \exp \left[ -\left( \kappa +\gamma \right) t/2\right] ,
\end{equation}%
where the subscript \textquotedblleft $\mathrm{e}$\textquotedblright\ refers
to the initial and final excited states. The exponential factor $\left(
\kappa +\gamma \right) /2$ can be considered as an effective decay rate. In
the RWA case, the qubit energy is splitting to $\Delta \pm g/2.$

While in our results beyond the RWA, the qubit energy splitting depends on
the qubit environment. Assume $\lambda =0.1~g,$ $\omega _{\mathrm{cav}%
}=100~g.$ If the qubit in the low-frequency bath with $\omega _{\mathrm{low}%
}=10~g$ and $\alpha _{\mathrm{low}}=10^{-4}$, the qubit energy is splitting
to $\Delta -0.4786~g$ and $\Delta +0.5011~g.$ While, if the qubit in the
Ohmic bath with $\omega _{\mathrm{Ohm}}=10^{3}~g$ and $\alpha _{\mathrm{Ohm}%
}=10^{-4},$ the qubit energy level is splitting to $\Delta -0.5018~g$ and $%
\Delta +0.4782~g.$

Figure \ref{Fig22} shows the probability $P_{\mathrm{e}}(t)$ for the qubit
to be in the excited state in the region $0<t<\pi $. When the qubit and the
cavity mode is resonant, the qubit decay with the measurements, whose
interval between successive measurements is $\tau =0.1~g^{-1},$ is slowed
down compared to the case without measurement (the interval $\tau $ extends
to infinite), which means QZE. While tune the cavity mode to $\omega _{%
\mathrm{cav}}=80~g$ and fix the energy level spacing of the qubit $\Delta
=100~g$, the decay with the measurements ($\tau =0.1~g^{-1}$) is speeded up
contrast to the case without the measurements, which means AZE.

\section{The effective decay rate of a qubit in a cavity with successive
measurements}

In the following, we will solve the Eq.~(\ref{e7}) iteratively and obtain
the effective decay rate with successive measurement \cite{pra-54-R3750,A10}%
. When the interval between measurements is sufficiently short, the
evolution of the qubit after measurements can be approximately expressed by
an exponential form. So the discussion in \cite{nature-405-546} can be
extended to damped oscillations. Namely, if the exponential factor is larger
or smaller than the effective decay rate $\left( \kappa +\gamma \right) /2,$
then the measurements reduce or enhance the decay rate. After the first
iteration, Eq.~(\ref{e7}) is solved as%
\begin{equation}
\widetilde{\chi }(t)\simeq 1-\int\limits_{0}^{t}(t-t^{^{\prime
}})\sum_{k,i}V_{k,i}^{2}\exp [-i(\omega _{k,i}-\eta \;\Delta )t^{^{\prime
}}]\;dt^{^{\prime }}.
\end{equation}%
For a small $t$, we can approximately write $\widetilde{\chi }(t)$ in an
exponential form:
\begin{widetext}
\begin{eqnarray}
\widetilde{\chi }(t) &=&\exp \left[ -\int\limits_{0}^{t}(t-t^{^{\prime
}})\sum_{k,i}V_{k,i}^{2}\exp \left[ -i(\omega _{k,i}-\eta \;\Delta
)t^{^{\prime }}\right] dt^{^{\prime }}\right]   \notag \\
&=&\exp \left\{ -t\left[ -\frac{1}{t}\sum_{k,i}V_{k,i}^{2}\frac{\exp \left[
-i\left( \omega _{k,i}-\eta \;\Delta \right) t\right] -1+i\left( \omega
_{k,i}-\eta \;\Delta \right) t}{\left( \omega _{k,i}-\eta \;\Delta \right)
^{2}}\right] \right\}   \notag \\
&=&\exp \left\{ -t\left[ \sum_{k,i}V_{k,i}^{2}\left( \frac{2\sin \left(
\frac{\omega _{k,i}-\eta \;\Delta }{2}t\right) ^{2}}{t\left( \omega
_{k,i}-\eta \;\Delta \right) ^{2}}-i\frac{\left( \omega _{k,i}-\eta \;\Delta
\right) t-\sin \left[ \left( \omega _{k,i}-\eta \;\Delta \right) t\right] }{%
t\left( \omega _{k,i}-\eta \;\Delta \right) ^{2}}\right) \right] \right\} .
\end{eqnarray}
\end{widetext}Note that\textit{\ only when }$\tau \ll g^{-1},$\textit{\ the
qubit evolution can be approximately described as an exponential decay }\cite%
{pra-40-5516,pra-54-R3750,pra-81-062131}, which has been reflected in
Fig.~2. Assume now that the instantaneously-ideal projection measurement is
performed periodically, separated by time intervals $\tau $. For a single
measurement, the probability amplitude of the qubit maintaining in the
initial state is $\widetilde{\chi }(t=\tau ).$ After a sufficiently large
number of measurements, the survival probability of the initial state
becomes
\begin{equation}
P _{\mathrm{e}}(t=n\tau )\;=\;\left\vert \widetilde{\chi }(t=n\tau
)\right\vert ^{2}\;=\;\exp [-\gamma (\tau )t].
\end{equation}%
And the exponential decay constant $\gamma (\tau )$ is obtained
\begin{eqnarray}
\gamma (\tau ) &=&2\pi \int_{0}^{\infty }d\omega \sum_{k,i}V_{k,i}^{2}\;%
\frac{2\sin ^{2}(\frac{\eta \;\Delta -\omega }{2}\tau )}{\pi (\eta \;\Delta
-\omega )^{2}\tau }  \notag \\
&=&2\pi \int_{0}^{\infty }d\omega J(\omega )f(\omega )F(\omega -\eta
\;\Delta ,\tau ),  \label{E43}
\end{eqnarray}%
where%
\begin{equation}
f(\omega )=\left( 1-\frac{\omega -\eta \;\Delta }{\omega +\eta \;\Delta }%
\right) ^{2},  \label{ef}
\end{equation}%
\begin{eqnarray}
J(\omega ) &=&\sum_{k}\left[ f_{k}^{2}\delta (\omega -\omega
_{k,1})+g_{k}^{2}\delta (\omega -\omega _{k,2})\right] \\
&=&J_{\mathrm{cav}}(\omega )+J_{\mathrm{qb}}(\omega ),  \label{e44}
\end{eqnarray}%
and%
\begin{equation}
F(\omega -\eta \;\Delta ,\tau )=\frac{2\sin ^{2}\left[ \left( \eta \;\Delta
-\omega \right) \tau /2\right] }{\pi (\eta \;\Delta -\omega )^{2}\tau }.
\label{e45}
\end{equation}%
In Eq.~(\ref{e44}), $J(\omega )$ is the entire interacting density spectrum
with $J_{\mathrm{cav}}(\omega )$ from the cavity bath and $J_{\mathrm{qb}%
}(\omega )$ the qubit's intrinsic bath. The function $F(\omega -\eta
\;\Delta ,\tau )$ comes from the projection measurements and can be called a
modulating function of the measurements.

The decay rate $\gamma (\tau )$, in Eq.~(\ref{E43}),$\ $depends on the
renormalization factor $\eta $ and $f(\omega )$ in Eq.~(\ref{ef}), which are
mainly from the anti-rotating terms. If we use the RWA, $\eta =1$ and $%
f(\omega )=1,$ which is consistent with the case of weak interaction.
Therefore, our results can apply to not only the weak coupling case, but
also to the case of \textit{strong coupling} between the qubit and the
environment. Furthermore, since the function $F(\omega -\Delta ,\tau )$
becomes $\delta (\omega -\eta \;\Delta )$ in the long-time limit, we obtain
the effective decay rate under the Weisskopf-Wigner approximation%
\begin{equation}
\gamma _{0}=\gamma (\tau \rightarrow \infty )=2\pi J(\eta \;\Delta ).
\end{equation}%
The normalized decay rate, which characterizes the QZE and the AZE, is
determined by
\begin{equation}
\frac{\gamma (\tau )}{\gamma _{0}}=\frac{\int_{0}^{\infty }d\omega J(\omega
)f(\omega )F(\omega -\eta \;\Delta ,\tau )}{J(\eta \;\Delta )}.  \label{E45}
\end{equation}%
For a finite time $\tau ,$ and when $\gamma (\tau )/\gamma _{0}<1$ holds, we
have the QZE, i.e., measurements hinder the decay. However, when $\gamma
(\tau )/\gamma _{0}>1,$ this implies the AZE, i.e., measurements enhance the
decay.

To see the contribution of each bath to the decay rate, Eq.~(\ref{E45}) can
be reexpressed as%
\begin{eqnarray}
&&\frac{\gamma (\tau )}{\gamma _{0}}  \notag \\
&=&\frac{J_{\mathrm{cav}}(\eta \;\Delta )}{J(\eta \;\Delta )}\frac{%
\int_{0}^{\infty }d\omega J_{\mathrm{cav}}(\omega )f(\omega )F(\omega -\eta
\;\Delta ,\tau )}{J_{\mathrm{cav}}(\eta \;\Delta )}  \notag \\
&&+\frac{J_{\mathrm{qu}}(\eta \;\Delta )}{J(\eta \;\Delta )}\frac{%
\int_{0}^{\infty }d\omega J_{\mathrm{qu}}(\omega )f(\omega )F(\omega -\eta
\;\Delta ,\tau )}{J_{\mathrm{qu}}(\eta \;\Delta )}.  \label{E46}
\end{eqnarray}%
From this Eq.~(\ref{E46}), we see that the normalized decay rate due to the
two baths is combined by the normalized decay rate from each bath by the
weights $J_{\mathrm{qb}}(\eta \;\Delta )/J(\eta \;\Delta )$ and $J_{\mathrm{%
cav}}(\eta \;\Delta )/J(\eta \;\Delta ),$ respectively.

\section{Results and Discussion}

In this section, we will show the normalized decay rate of the qubit-cavity
system in three cases: $\left( i\right) $ only the cavity bath, $\left(
ii\right) $ both the cavity bath and the low-frequency qubit spontaneous
dissipation bath coexist, as well as both the cavity bath and the Ohmic
qubit's intrinsic bath coexistence. According to the experiment \cite%
{nature-431-162}, we consider the qubit weakly coupled to the qubit
intrinsic bath with coupling constants $\alpha _{\mathrm{Ohm}}=10^{-4}$ and $%
\alpha _{\mathrm{low}}=10^{-4}.$ The quality factor $Q$ of the cavity is
assumed in the range of $2\times 10^{2}\sim 10^{4}$.

\subsection{Only cavity bath}

Let us first consider the case of a qubit only in a cavity bath. For
example, when the qubit-cavity interaction $g\gg \alpha _{\mathrm{low}%
}\Delta ,$ or $g\gg \alpha _{\mathrm{Ohm}}\Delta $, which has been realized
in a superconducting qubit coupled to a transmission line cavity \cite%
{nature-458-178,add8}. For such strong coupling between the qubit and
cavity, the normalized decay rate mainly depends on the cavity bath. Then in
this case, the decay rate can be approximately written as\
\begin{equation}
\gamma (\tau )=2\pi \int_{0}^{\infty }d\omega \;J_{\mathrm{cav}}(\omega
)\;f(\omega )\;F(\omega -\eta \;\Delta ,\tau ).
\end{equation}%
From the normalized decay rate in Eq.~(\ref{E45}), we see that the
qubit-cavity coupling strength $g$ is in both, the numerator and
denominator, so it cancels out. Therefore, the normalized decay rate $\gamma
(\tau )$ is independent of the qubit-cavity coupling strength $g.$ However
we still note that\textit{\ only in the case when }$\tau \ll g^{-1},$\textit{%
\ the qubit evolution can be approximately described by an exponential decay.%
} This means that if there is a strong qubit-cavity coupling $g=0.1\Delta ,$
the measurement interval becomes $\tau \ll g^{-1}\sim 10\Delta ^{-1}.$ When
the qubit-cavity is not so strong, $g=10^{-2}\Delta ,$ the measurement
interval could be $\tau \ll g^{-1}\sim 10^{2}\Delta ^{-1}.$

\begin{figure}[tbp]
\includegraphics[width=3.2in,clip]{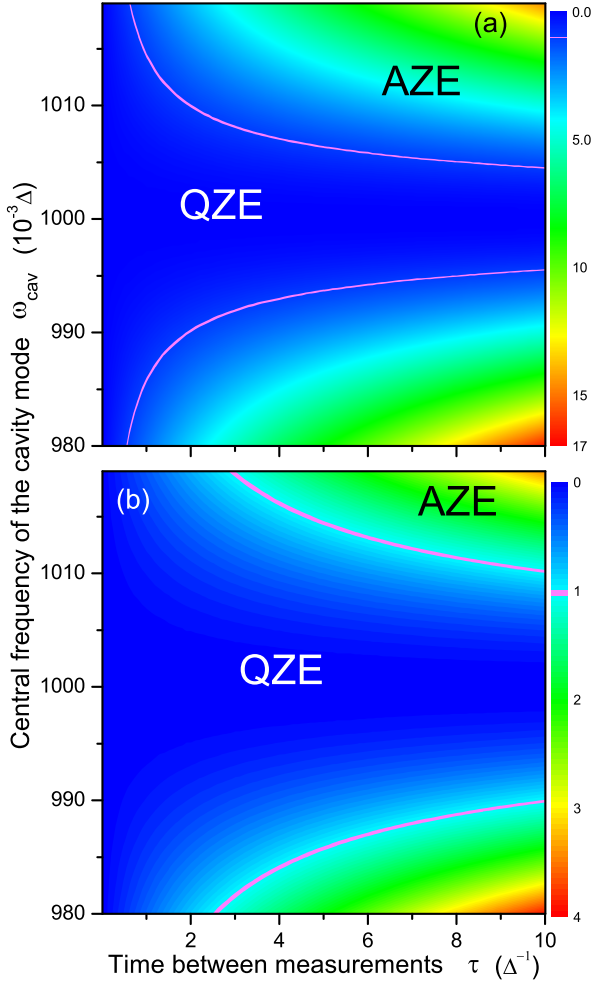}
\caption{(Color online) Contour plots of the normalized decay rate $\protect%
\gamma (\protect\tau )/\protect\gamma _{0}$ of the qubit only in the cavity
bath, versus the time interval $\protect\tau $ between successive
measurements, and the central frequency $\protect\omega _{\mathrm{cav}}$ of
the cavity mode. (a) The width of the cavity frequency is $\protect\lambda %
=10^{-4}\Delta$, and accordingly the cavity quality factor $Q = 10^{4}.$ (b)
The width of the cavity frequency $\protect\lambda =5\times 10^{-3}\Delta$,
corresponding to the cavity quality factor $Q = 2\times 10^{3}$. The region $%
1 \leq \protect\gamma (\protect\tau )/\protect\gamma _{0}\leq 1.05$ is shown
as light magenta. The QZE region corresponds to $\protect\gamma (\protect%
\tau )/\protect\gamma _{0} <1$. The AZE region covers the rest, when $%
\protect\gamma (\protect\tau )/\protect\gamma _{0} >1$. Evidently, a
transition from the QZE to the AZE is observed by varying the central
frequency of the cavity mode at finite $\protect\tau$ ($\protect\tau >0.6
\Delta^{-1}$ when $Q = 10^{4}$, and $\protect\tau >2.6 \Delta^{-1}$ when $Q
= 2\times 10^{3}$).}
\label{Fig2}
\end{figure}

Figure~\ref{Fig2} displays the normalized decay rate as a function of the
measurement interval $\tau $ and the cavity central frequency $\omega _{%
\mathrm{cav}}.$ Figures~\ref{Fig2}(a) and (b) correspond to two quality
factors of the cavity: $Q=10^{4}$ and $Q=2\times 10^{3}$, respectively. We
can see that in the limit when $\tau \rightarrow 0,$ only the QZE occurs. \
For a finite interval, the normalized decay rate of the qubit exhibits a
transition from the QZE to the AZE, by modulating the central frequency of
the cavity mode $\omega _{\mathrm{cav}}$ in and off resonance with the qubit
energy level spacing $\Delta $. \textit{The variation should be useful to
distinguish the QZE and the AZE.}

Let us now estimate the condition for the transition between the QZE and the
AZE. From Fig.~\ref{Fig2}(a), the crossover from QZE to AZE, by varying the
cavity frequency, appears only for the measurement interval $\tau >0.6\Delta
^{-1}$. Using the condition $\tau \ll g^{-1}$, we obtain the qubit-cavity
coupling strength $g\ll 1.7\Delta .$ Similarly, for the cavity quality
factor $Q=2\times 10^{3},$ we obtain the qubit-cavity coupling strength $%
g\ll 0.38\Delta .$

\begin{figure}[tbp]
\includegraphics[width=3.2in,clip]{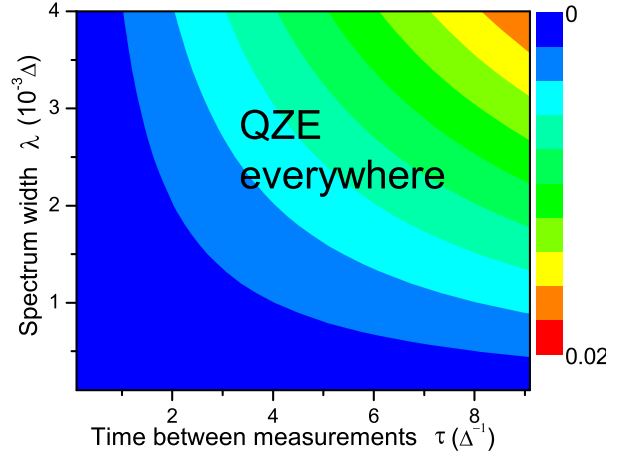}
\caption{(Color online) Contour plots of the normalized effective decay rate
$\protect\gamma (\protect\tau )/\protect\gamma _{0}$ of the qubit in the
resonant case $\Delta =\protect\omega _{\mathrm{cav}}$.}
\label{Fig3}
\end{figure}

In Fig.~\ref{Fig3}, we plot the normalized decay rate with the cavity
frequency in resonance with the qubit, $\omega _{\mathrm{cav}}=\Delta $,
versus the time interval $\tau $ between successive measurements, and the
cavity spectral width $\lambda $. It is obvious that only the QZE exists in
the resonant case. The normalized decay rate $\gamma (\tau )/\gamma _{0}$
becomes smaller as the cavity spectral width $\lambda$ decreases. This
indicates that the transition from the QZE to the AZE becomes sharper as the
cavity spectral width $\lambda$ reduces.

To better understand the transition from the QZE to the AZE, we discuss the
results in two regimes of near-resonance (including on-resonance) and
off-resonance (the central frequency of the cavity mode\ higher and lower
than the qubit spacing $\Delta $):

1. In the case of on-resonance $\Delta =\omega _{\mathrm{cav}},$ and
near-resonance $\left\vert \Delta -\omega _{\mathrm{cav}}\right\vert
<\lambda $, without measurements, the effective decay rate of the qubit is
given by $J(\Delta ).$ Moreover, the qubit is resonant with the cavity mode,
$\Delta =\omega _{\mathrm{cav}}$. Note that $\omega _{\mathrm{cav}}$ is the
peak of the density spectrum of the cavity-bath, where the probability of
energy transfer from the qubit to the cavity bath is maximum. In this case,
the qubit strongly decays in its evolution. Every measurement to project the
qubit on the initial state protects the qubit from decay, i.e., protects the
qubit from exchanging energy with the cavity. From Eq.~(\ref{e45}), the
modulating function $F(\omega -\eta \Delta ,\tau )$ of the measurements is a
periodically oscillating function versus energy $\omega $ for a fixed time
interval $\tau $. Moreover, its integral over all energies is $1$. Thus we
consider each oscillator peak as a decay channel induced by measurements.
Without measurements, $F(\omega -\eta \;\Delta ,\tau )$ becomes $\delta
(\omega -\eta \;\Delta ,\tau )$. Only one channel $\omega =\eta \;\Delta $
exists. With measurements, more channels will appear, but the probability of
qubit-energy-decay via every channel decreased to less than 1. Among these
channels, the largest one is still $\omega =\eta \;\Delta $, which is less
than the non-measurement one. Therefore, the superposition of the density
spectrum function $J_{\mathrm{cav}}(\omega )$ of the cavity-bath and the
modulating function $F(\omega -\eta \;\Delta ,\tau )$ of the measurements
reduces the effective decay rate and protects the qubit energy from leaking
to the cavity-bath when the qubit is resonantly coupled to the cavity.

2. For the case of off-resonance $\left\vert \Delta -\omega _{\mathrm{cav}%
}\right\vert >\lambda ,$ especially for the large-detuning limit $\left\vert
\Delta -\omega _{\mathrm{cav}}\right\vert \gg \lambda ,$ the effective
interaction between the qubit and the cavity becomes very weak. For example,
the ratio of the effective decay rate in $\omega _{\mathrm{cav}}=0.98\Delta $
(or $\omega _{\mathrm{cav}}=1.02\Delta $) to $\omega _{\mathrm{cav}}=\Delta $
is about $2\times 10^{-5}.$ In most quantum optics papers, large-detuning
means that the qubit is free from decay. Thus, the probability of the qubit
maintaining its initial state is close to 1. After introducing frequent
measurements, the qubit suffers from AZE, i.e., measurements enhance the
decay, which is opposite to the on-resonant case. The reason for this also
comes from the modulating function of the measurements $F(\omega -\eta
\;\Delta ,\tau ),$ a periodic oscillation function of the energy. As long as
one of the oscillation peaks of $F(\omega -\eta \;\Delta ,\tau )$ is located
in the effective region of the density spectrum $J(\omega )$, especially the
half-width of the maximum, the product of these two functions will lead to
an enhancement of the effective decay rate. In other words, the\ periodic
oscillations of $F(\omega -\eta \;\Delta ,\tau )$ connect the qubit energy
with the density spectrum of the cavity bath and open the decay channels of
the qubit\ energy to the cavity bath. From this point of view, the
measurements act as a new decay element, besides the cavity and the qubit
intrinsic bath. The AZE becomes more obvious as the detuning increases.

In the above discussion, we have investigated the qubit decay dynamics
subject to measurements mainly induced by the cavity bath. Also, we have
studied the qubit decay dynamics subject to measurements due to either the
low-frequency qubit spontaneous dissipation bath or the Ohmic qubit
intrinsic bath in Ref.~[\onlinecite{arkiv}]. In the next two subsections, we
will show the normalized effective decay rate of the qubit in the presence
of both the cavity bath and the qubit's intrinsic bath.

\subsection{Coexistence of the cavity bath and the low-frequency qubit
intrinsic bath}

\begin{figure}[tbp]
\includegraphics[width=3.2in,clip]{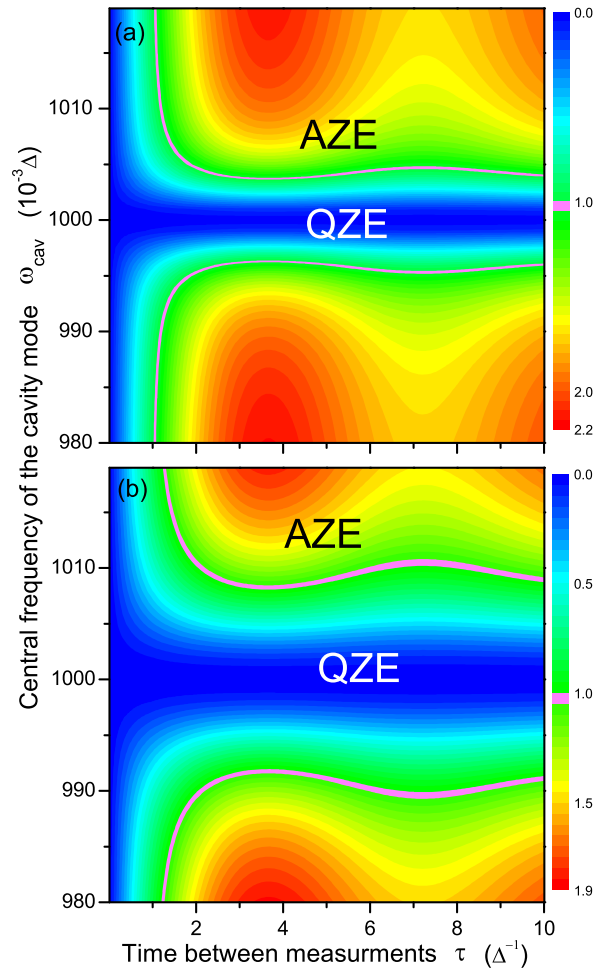}
\caption{(Color online) Contour plots of the normalized effective decay rate
$\protect\gamma (\protect\tau )/\protect\gamma _{0}$ in the presence of both
the cavity bath and the low-frequency qubit's intrinsic bath. The
interaction strength $\protect\alpha _{\mathrm{low}}=10^{-4}$, between the
qubit and qubit's intrinsic bath, and the qubit-cavity coupling $%
g=10^{-2}\Delta $. (a) Results for the cavity quality factor $Q = 10^{4}.$
(b) Results for the cavity quality factor $Q = 2\times 10^{3}.$ The region $%
1 \leq \protect\gamma (\protect\tau )/\protect\gamma _{0}\leq 1.05 $ is
shown as light magenta. The QZE region corresponds to $\protect\gamma (%
\protect\tau )/\protect\gamma _{0} <1$. The AZE region covers the rest, when
$\protect\gamma (\protect\tau )/\protect\gamma _{0} >1$. Evidently, a
transition from the \ QZE to the AZE is observed by varying the central
frequency $\protect\omega _{\mathrm{cav}}$ of the cavity mode at finite $%
\protect\tau$.}
\label{Fig4}
\end{figure}

\begin{figure}[tbp]
\includegraphics[width=3.2in,clip]{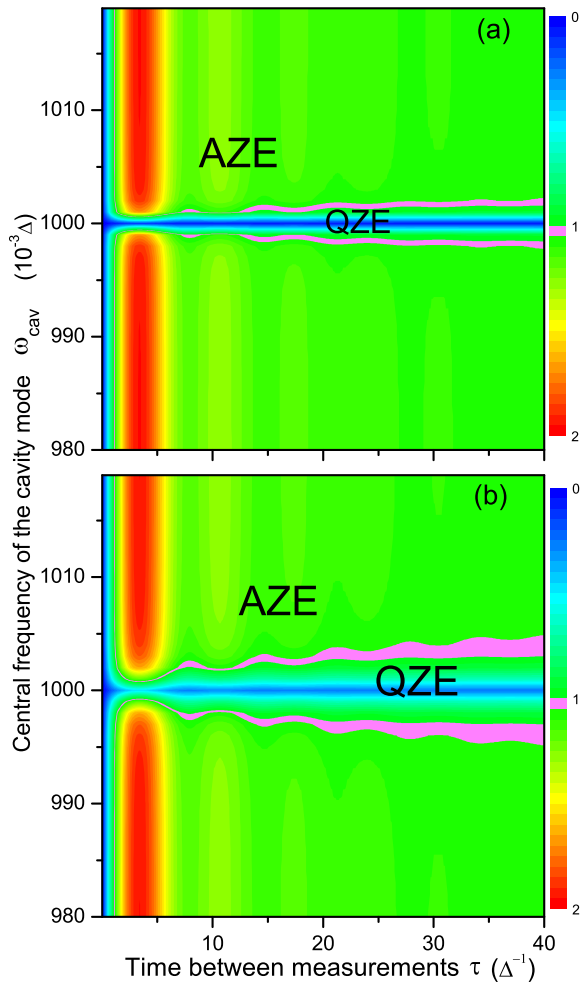}
\caption{(Color online) The qubit-cavity coupling is $g=10^{-3}\Delta $. The
other caption is the same as Fig.~\protect\ref{Fig4}.}
\label{Fig5}
\end{figure}

In Figs.~\ref{Fig4} and~\ref{Fig5}, we plot the normalized effective decay
rate, when the cavity bath and the low-frequency qubit spontaneous
dissipation bath coexist, versus the time interval $\tau $ between
measurements in the regime of strong ($g=10^{-2}\Delta $) and weak ($%
g=10^{-3}\Delta $) cavity-qubit coupling with the cavity central frequency
around the qubit energy-level-spacing $\Delta .$ Figures~\ref{Fig4}(a) and
(b) correspond to the cavity quality factor $Q=10^{4}$ and $Q=2\times 10^{3}$%
, respectively$.$ From Fig.~\ref{Fig4}, we see that for a strong
qubit-cavity coupling, by modulating the cavity central frequency from
in-resonance to off-resonance with the qubit energy-level-spacing, the
normalized effective decay rate grows and becomes larger than $1,$ which
clearly displays the transition from the QZE to the AZE. Comparing Figs.~\ref%
{Fig4}(a) and (b), as the width $\lambda $ of the cavity frequency decreases
(or the quality factor $Q=\omega _{\mathrm{cav}}/\lambda$ increases), the
region of the cavity frequency for the QZE becomes narrower. For example,
Table I presents the normalized effective decay rate $\gamma (\tau )/\gamma
_{0}$ for two quality factors $Q$ and different central frequencies $\omega
_{\mathrm{cav}}$ of the cavity, when $\tau =5\;\Delta ^{-1}$.
\begin{table*}[tbp]
\caption{The normalized effective decay rate $\protect\gamma (\protect\tau )/%
\protect\gamma _{0}$ of the qubit for two quality factors $Q$ when $\protect%
\tau =5\;\Delta ^{-1}$, in the presence of both the cavity bath and the
low-frequency qubit's intrinsic bath.}
\begin{center}
\begin{tabular}{|c||c|c|c|c|c|c|c|}
\hline
\multirow{2}{*} { Cavity quality factor } & \multicolumn{7}{|c|}{Central
frequency of the cavity} \\
\hhline{~-------} & \; $0.98\Delta$\; & \;$\; 0.99\Delta$ \; & \; $%
0.999\Delta$ \; & \; \; $\Delta$ \; \; & \; $1.001\Delta$ \; & \; $%
1.01\Delta $ \; & \; $1.02\Delta$\; \\ \hline\hline
$Q=10^{4}$ & 1.994 & 1.784 & 0.122 & 0.001 & 0.122 & 1.777 & 1.979 \\ \hline
$Q=2\times 10^{3}$ & 1.727 & 1.149 & 0.032 & 0.006 & 0.032 & 1.145 & 1.714
\\ \hline
\end{tabular}%
\end{center}
\end{table*}

For weak qubit-cavity coupling in Fig.~\ref{Fig5}, only in the short-time
regime (about $2\Delta ^{-1}<\tau <6\Delta ^{-1}$), the normalized effective
decay rate of the qubit shows obviously the transition from the QZE to the
AZE. When the measurement interval $\tau $ increases to $\tau >10\Delta
^{-1} $, although the transition still exists, $\gamma (\tau )/\gamma _{0}$
for the AZE is slightly larger than 1, which is mainly in the region between
$1.0$ and $1.1.$

In Figs.~\ref{Fig4} and~\ref{Fig5}, there appear distinct
oscillations in the qubit's QZE-AZE transition processes. From the
results of Ref.~[\onlinecite{arkiv}], we have known that the AZE
always occurs for a qubit in the low-frequency bath. While, for a
qubit in the cavity bath, the transition from QZE to AZE takes place
by varying the cavity frequency. These oscillations in
Figs.~\ref{Fig4}-\ref{Fig5} come from the different impacts of the
cavity-bath and the low-frequency bath on the qubit's measurement
dynamics.

\subsection{Coexistence of the cavity bath and the Ohmic qubit spontaneous
dissipation bath}

\begin{figure}[tbp]
\includegraphics[width=3.2in,clip]{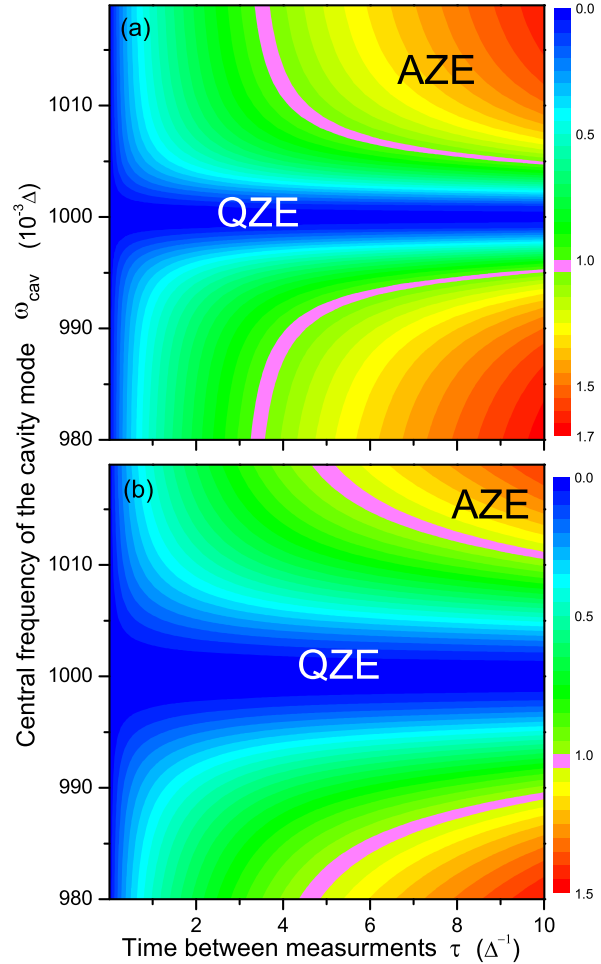}
\caption{(Color online) Contour plots of the normalized effective decay rate
$\protect\gamma (\protect\tau )/\protect\gamma _{0}$ in the presence of
both: the cavity bath and the Ohmic qubit's intrinsic bath. The interaction
strength $\protect\alpha _{\mathrm{Ohm}}=10^{-4}$, between the qubit and the
qubit's intrinsic bath. Also the qubit-cavity coupling $g=10^{-2}\Delta $.
(a) The cavity quality factor of the cavity $Q = 10^{4}.$ (b) The cavity
quality factor $Q = 2\times 10^{3}.$ The region $1 \leq \protect\gamma (%
\protect\tau )/\protect\gamma _{0}\leq 1.05$ is shown as light magenta. The
QZE region corresponds to $\protect\gamma (\protect\tau )/\protect\gamma %
_{0} <1$. The AZE region is the rest, when $\protect\gamma (\protect\tau )/%
\protect\gamma _{0} >1$. Evidently, a transition from the \ QZE to the AZE
is observed by varying the central frequency $\protect\omega _{\mathrm{cav}}$
of the cavity mode at finite $\protect\tau$.}
\label{Fig6}
\end{figure}

\begin{figure}[tbp]
\includegraphics[width=3.2in,clip]{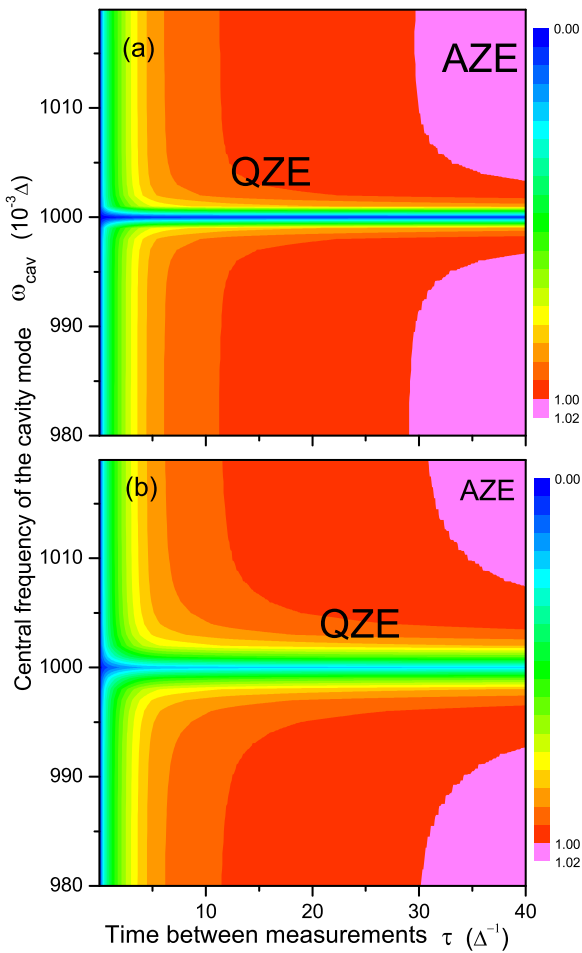}
\caption{(Color online) The qubit-cavity coupling $g=10^{-3}\Delta $. The
other caption is the same as Fig.~\protect\ref{Fig6}}
\label{Fig7}
\end{figure}

In Figs.~\ref{Fig6} and \ref{Fig7}, we show the normalized effective decay
rate $\gamma (\tau )/\gamma _{0}$ in the presence of both the Ohmic
intrinsic bath and cavity bath. Comparing Figs.~\ref{Fig6} with~\ref{Fig4},
we find that for the strong qubit-cavity coupling $g=10^{-2}\Delta ,$ the
time interval $\tau $ for the QZE increases in the short-time region. In the
long-time region, the features of Figs.~\ref{Fig4} and \ref{Fig6} are almost
identical. From Fig.~\ref{Fig7} we can see that in the short-time region, ($%
0<\tau <30\Delta ^{-1})$, only the QZE exists, regardless of the central
frequency of cavity. For $\tau >30\Delta ^{-1},$ the normalized effective
decay rate $\gamma /\gamma _{0}$ for the AZE is in the small region of $%
1.0\sim 1.02$, which is not conducive to observe the transition from the QZE
to the AZE.

\section{Summary}

We investigated the QZE and AZE of a qubit in a cavity when both the cavity
bath and the qubit's intrinsic bath (either low-frequency or Ohmic bath) are
simultaneously present. We find that in the case of strong qubit-cavity
coupling, modulating the cavity central frequency from on-resonance ($\omega
_{\mathrm{cav}}=\Delta )$ to off-resonance ($\omega _{\mathrm{cav}}$ larger
or smaller than $\Delta )$ with the qubit energy-level-spacing, the
transition from the QZE to the AZE occurs. Thus, our results provide a
proposal to observe the QZE and the AZE in the qubit-cavity system.

{\noindent {\large \textbf{Acknowledgements}}}

\vskip0.5cm

We thank A.~G.~ Kofman for comments on the manuscript. FN acknowledges
partial support from DARPA, AFOSR, the National Security Agency (NSA),
Laboratory for Physical Sciences (LPS), Army Research Office (USARO),
National Science Foundation (NSF) under Grant No. 0726909, JSPS-RFBR under
Contract No. 09-02-92114, MEXT Kakenhi on Quantum Cybernetics, and FIRST
(Funding Program for Innovative R\&D on S\&T). X.-F. Cao acknowledges
support from the National Natural Science Foundation of China under Grant
No. 10904126 and Fujian Province Natural Science Foundation under Grant No.
2009J05014.

--------------------

\end{document}